# Broadband Continuous Frequency Tuning in Non-Hermitian Laser Arrays Enabled by Mode-Switching Boundary Topology

Chuanfeng Yan[1,2,†], Cheng Tan[1,2,†], Kai Wang[1], Hongzhou Bai[1,2], Shanhai Gao[1,2], Lianghua Gan[1], Yueheng Zhang[3,*], Qijie Wang[4], Gangyi Xu[1,*]

1. *National Key Laboratory of Infrared Detection Technologies, Shanghai Institute of Technical Physics, Chinese Academy of Sciences, Shanghai 200083, China*
2. *University of Chinese Academy of Sciences, Beijing 100049, China*
3. *Key Laboratory of Artificial Structures and Quantum Control, School of Physics and Astronomy, Shanghai Jiao Tong University, Shanghai 200240, China*
4. *Division of Physics and Applied Physics, School of Physical and Mathematical Sciences, Nanyang Technological University, 21 Nanyang Link, Singapore 637371, Singapore*

\* Corresponding authors: *yuehzhang@sjtu.edu.cn, gangyi.xu@mail.sitp.ac.cn*

† These authors contribute equally to this work.

**Abstract**

Broadband and continuous frequency tuning is central to the versatility of semiconductor lasers, yet existing approaches typically rely on external moving components, limiting scalability and integration. Here we demonstrate broadband continuous frequency tuning in a non-Hermitian laser array achieved solely by controlling the pump currents. We show that in two coupled sub-lasers with frequency detuning ($\Delta\omega$) and relative loss ($\Delta\alpha$), a mode-switching boundary emerges in the ($\Delta\omega$, $\Delta\alpha$) parameter space, shaping the frequency landscape of the lower-loss supermode. The topology of this boundary comprises pseudo-symmetric (PS) and pseudo-symmetry-broken (PSB) branches connected at an exceptional point (EP). When tuning trajectories cross the PS branch, frequency tuning is discontinuous, whereas trajectories cross the PSB branch enable continuous tuning; trajectories through the EP yield the maximum continuous tuning range. Experiments using two coupled terahertz quantum cascade lasers demonstrate continuous tuning over 10 GHz, enabled by arbitrarily many combinations of pump currents. Extending this approach to a multi-element array further expands the continuous tuning range to 163 GHz. These results establish a general route to broadband continuous tuning in moving-part-free semiconductor lasers and highlight the potential for dynamic eigenvalue engineering in non-Hermitian photonics and beyond.

# Introduction

Frequency tuning is a defining capability of semiconductor lasers, underpinning their versatility in spectroscopy, sensing, communications, and signal processing. While narrow or discrete tuning can be achieved through thermal or carrier-induced refractive index modulation[1-3], realizing broadband continuous frequency tuning remains a longstanding challenge. Existing approaches for wide tuning typically rely on external cavities with mechanically moving components to modify the effective cavity length[4-6], refractive index[7], or facet reflectance spectrum[8-10], at the cost of increased system size and complexity, and limited scalability for integrated applications. Developing a compact, monolithic, and mechanically static scheme for broadband continuous frequency tuning therefore remains a crucial goal in semiconductor laser science.

Non-Hermitian optical systems, characterized by complex eigenvalues and non-orthogonal eigenvectors[11,12], offer new opportunities for frequency control beyond conventional Hermitian designs. In such systems, the eigenfrequencies depend not only on the intrinsic properties of individual resonators but also on gain–loss contrast and inter-resonator coupling. A prominent feature of non-Hermitian systems is the exceptional point (EP), where both eigenvalues and eigenvectors coalesce, giving rise to branch-point singularities in parameter space and a variety of counterintuitive phenomena[13-22]. These effects have been explored extensively in optics and photonics, enabling unconventional functionalities[23-25] such as unidirectional invisibility[26,27], lasing self-termination[28,29], light trapping[30], and enhanced sensitivity[31-33]. Despite this progress, however, non-Hermitian photonics has not yet provided a general route to broadband continuous eigenfrequency tuning, particularly in laser systems. Prior studies on coupled microtoroidal whispering-gallery-mode resonators[34], photonic crystals[35,36], and vertical cavity lasers[37] have revealed dispersive coupling and eigenfrequency repulsion[38], often accompanied by mode-hopping or multimode lasing[34,36]. Nevertheless, achieving broadband continuous tuning in coupled-laser configurations has remained elusive.

Here, we demonstrate broadband continuous frequency tuning in a moving-part-free non-Hermitian laser array by exploiting the topology of the mode-switching boundary existing in the system's complex eigenvalue landscape. For two coupled lasers characterized by frequency detuning $\Delta\omega$ and relative loss $\Delta\alpha$, a mode-switching boundary emerges in the ($\Delta\omega$, $\Delta\alpha$) parameter space, separating regions in which different supermodes exhibit the lowest loss and thus dominate lasing. The topology of this boundary, which reflects the global connectivity of the complex eigenvalue surfaces, comprises pseudo-symmetric (PS) and pseudo-symmetry-broken (PSB) branches that meet at an EP. Here, PS and PSB branches are characterized by degeneracy in the imaginary and real parts of the supermode frequencies, respectively.

Frequency tuning is realized by varying the pump currents and thus the relative loss $\Delta\alpha$ of the sub-lasers, which drives the system along trajectories in the ($\Delta\omega$, $\Delta\alpha$) parameter space that necessarily cross the mode-switching boundary. The resulting frequency response is governed by the relation between the tuning trajectory and the EP. Trajectories above the EP (crossing the PS branch) lead to discontinuous frequency hopping, whereas trajectories below the EP (crossing the PSB branch) support continuous frequency tuning. The trajectories passing through the EP yields the maximum continuous tuning range. This trajectory-dependent tuning behavior is a direct manifestation of the topology of the mode-switching boundary, rather than of specific material parameters or implementation details.

We experimentally validate this principle using terahertz quantum cascade lasers (THz QCLs), a platform where broadband continuous tuning has remained particularly challenging[5-7,10,39-43]. In a two-laser system, we demonstrate continuous single-mode tuning over 10 GHz, enabled by arbitrarily many combinations of pump currents. Extending this approach to a multi-element laser array further expands the continuous tuning range to 163 GHz, comparable to external-cavity systems while maintaining a compact, monolithic, and mechanically static architecture.

# Theory

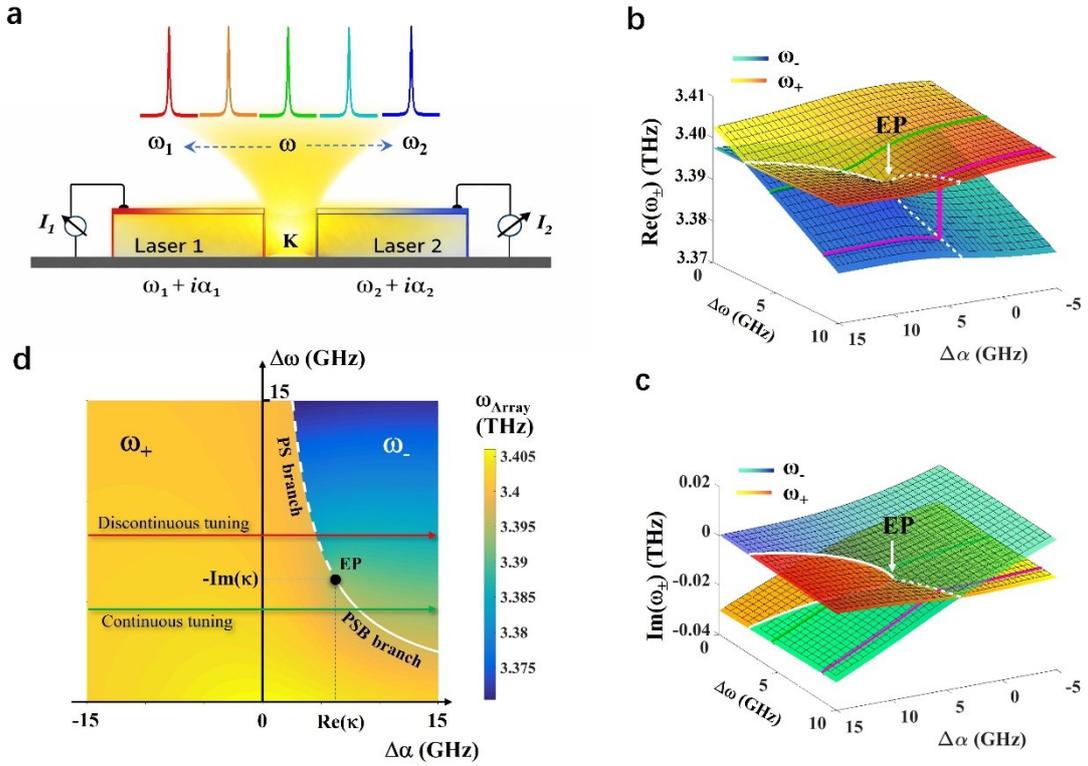

**Figure 1 | Complex coupling and frequency tuning in a non-Hermitian dual-laser system. a,** Schematic of two coupled sub-lasers with pump currents $I_i$, solitary frequencies $\omega_i$ and losses $\alpha_i$ (i = 1, 2), and a complex coupling coefficient $\kappa$. The frequency detuning and relative loss are $\Delta\omega = (\omega_1-\omega_2)/2$ and $\Delta\alpha = (\alpha_1-\alpha_2)/2$. **b, c,** Real (**b**) and imaginary (**c**) parts of the complex eigenfrequencies $\omega_+$ and $\omega_-$ of the two supermodes plotted in the ($\Delta\omega$, $\Delta\alpha$) parameter space. Solid (dashed) white curves mark intersections where the real (imaginary) parts of $\omega_+$ and $\omega_-$ degenerate, while the imaginary (real) parts split. **d,** Real frequency of the lower-loss supermode $\omega_{Array}$ plotted in the ($\Delta\omega$, $\Delta\alpha$) parameter space. The mode-switching boundary, separating regions where $\omega_+$ or $\omega_-$ has lower loss, consists of a PSB branch (solid white curve) and a PS branch (dashed white curve), which meet at an EP. Frequency tuning is continuous when the tuning trajectory crosses the PSB branch (green line), but discontinuous when crossing the PS branch (red line). The meanings of the curves and trajectories are consistent across panels **b–d**. To visualize the model, in panels **b-d**, $\omega_1+i\alpha_1$ is set as 3.4+0i (THz), $\kappa$ equals to 6.0-6.3i (GHz), and the scan ranges of $\Delta\omega$ and $\Delta\alpha$ are marked in each panel.

We develop a general theoretical model for frequency tuning in a system of two complexly coupled lasers, which captures the essential topological structure of the mode-switching boundary underlying broadband continuous tuning. Figure 1a schematically illustrates such a two-level non-Hermitian system, where the uncoupled lasing frequencies of the two sub-lasers are $\omega_1$ and $\omega_2$ ($\omega_1 > \omega_2$), and their net loss rates (intrinsic loss minus material gain) are $\alpha_1$ and $\alpha_2$, respectively. The mutual coupling between the two sub-lasers is assumed to be reciprocal and described by a complex coupling coefficient $\kappa = Re(\kappa) + i\,Im(\kappa)$. Within temporal coupled-mode theory (TCMT) under linear gain, the complex eigenfrequencies of the system are[44]:

$$\omega_\pm = \omega_0 + i\alpha_0 \pm \sqrt{[Re(\kappa)^2 - Im(\kappa)^2 + \Delta\omega^2 - \Delta\alpha^2] + 2i[Re(\kappa)Im(\kappa) + \Delta\omega\Delta\alpha]} \quad (1)$$

where $\omega_0 = (\omega_1 + \omega_2)/2$ and $\alpha_0 = (\alpha_1 + \alpha_2)/2$ denote the average frequency and loss, while $\Delta\omega = (\omega_1 - \omega_2)/2$ and $\Delta\alpha = (\alpha_1 - \alpha_2)/2$ represent the frequency detuning and relative loss between the two sub-lasers, respectively.

The parameters $(\Delta\omega, \Delta\alpha)$ define a two-dimensional space in which the complex eigenfrequencies $\omega_\pm$ form two connected Riemann surfaces, as illustrated by their real and imaginary parts in Figs. 1b and 1c. Here, for visualization, κ is taken as constant, although this assumption is not essential. In these plots, solid white curves mark traces where the real parts of $\omega_\pm$ degenerate while the imaginary parts split, whereas dashed white curves indicate the opposite case. These two traces meet at an EP, where both the real and imaginary parts of $\omega_\pm$ simultaneously coalesce.

Lasing occurs in the supermode with the smaller imaginary part of the eigenfrequency. According to Eq. (1), synchronously varying $\alpha_1$ and $\alpha_2$—and hence the average loss $\alpha_0$—while keeping $\Delta\omega$ and $\Delta\alpha$ fixed allows the lower-loss supermode to reach the lasing threshold without changing its real frequency. The resultant lasing frequency ($\omega_{Array}$), corresponding to the real part of the lower-loss supermode, is plotted in Fig. 1d as a function of $(\Delta\omega, \Delta\alpha)$. Equation (1) indicates that the system operates in the $\omega_+$ supermode when $Re(\kappa)Im(\kappa) + \Delta\omega\Delta\alpha < 0$, and switches to $\omega_-$ otherwise. The mode-switching boundary defined by $Re(\kappa)Im(\kappa) = -\Delta\omega\Delta\alpha$, marked by the dashed and solid white curves in Fig. 1d, therefore separates regions where different supermodes dominate lasing and coincides with the intersections of the Riemann surfaces in Figs. 1b and 1c.

As explained in Supplementary Information S1, the topology of the mode-switching boundary comprises two distinct branches. For $\Delta\omega > |Im(\kappa)|$ (the dashed white curve in Fig. 1d), the imaginary parts of $\omega_\pm$ coincide while the real parts split, defining the PS branch. For $\Delta\omega < |Im(\kappa)|$ (the solid white curve), the real parts coincide while their imaginary parts split, corresponding to the PSB branch. These two branches meet at the EP, where $\Delta\omega = -Im(\kappa)$ and $\Delta\alpha = Re(\kappa)$ (or equivalently, $\Delta\omega = Im(\kappa)$ and $\Delta\alpha = -Re(\kappa)$, depending on the sign of $Im(\kappa)$), and both eigenfrequencies fully coalesce. The EP thus appears as a branch point on the mode-switching boundary.

In experiments, the frequencies $\omega_{1,2}$—and hence $\Delta\omega$—are fixed after fabrication, whereas the relative loss $\Delta\alpha$ can be continuously changed via the pump currents. As a result, the experimentally accessible tuning trajectories in the $(\Delta\omega, \Delta\alpha)$ parameter space are approximately parallel to the $\Delta\alpha$ axis and necessarily cross the mode-switching boundary. The nature of the frequency response is therefore governed by the relation between the tuning trajectories and the EP. As illustrated in Fig. 1d, trajectories above the EP traverse the PS branch and result in discontinuous frequency hopping, whereas trajectories below the EP traverse the PSB branch and support continuous frequency tuning.

We now focus on tuning trajectories below the EP. As detailed in Supplementary Information S1, the evolution of the lasing frequency $\omega_{Array}$ can be determined as $\Delta\alpha$ is swept from -∞ to +∞. Starting from large negative $\Delta\alpha$, the system operates on the $\omega_+$ supermode, with $\omega_{Array}$ initially close to $\omega_1$. As $\Delta\alpha$ increases, $\omega_{Array}$ rises and reaches its maximum value $\omega_0 + \sqrt{\text{Re}(\kappa)^2 + \Delta\omega^2}$ at $\Delta\alpha = \Delta\omega\,\text{Im}(\kappa)/\text{Re}(\kappa)$. With further increase of $\Delta\alpha$, $\omega_{Array}$ decreases and reaches $\omega_0$ at the mode-switching boundary, where the lasing mode switches from $\omega_+$ to $\omega_-$, with $\text{Re}(\omega_+) = \text{Re}(\omega_-) = \omega_0$. Upon further increasing $\Delta\alpha$, $\omega_{Array}$ decreases and asymptotically approaches $\omega_2$ as $\Delta\alpha \to +\infty$.

Throughout this process, provided that $\Delta\alpha$ is varied over a sufficiently large range, the lasing frequency $\omega_{Array}$ can be continuously tuned from near $\omega_1$ to near $\omega_2$, yielding a total tuning range of $\left(\Delta\omega + \sqrt{\text{Re}(\kappa)^2 + \Delta\omega^2}\right)$. Consequently, the most efficient tuning trajectory is the one crossing the EP, which produces the largest continuous tuning range for a given variation range of $\Delta\alpha$. Notably, the main tuning characteristics derived from our theoretical model remain valid even when $\kappa$ varies with $\Delta\omega$ and $\Delta\alpha$, as confirmed by the numerical simulations in the next section.

**Device design**

Guided by the general theory and its topological interpretation, we design two coupled THz QCLs that enable controlled exploration of tuning trajectories relative to the EP on the mode-switching boundary. The device architecture is engineered such that the experimentally accessible control parameters map directly onto the $(\Delta\omega, \Delta\alpha)$ space, allowing both continuous and discontinuous frequency tuning to be realized and systematically compared.

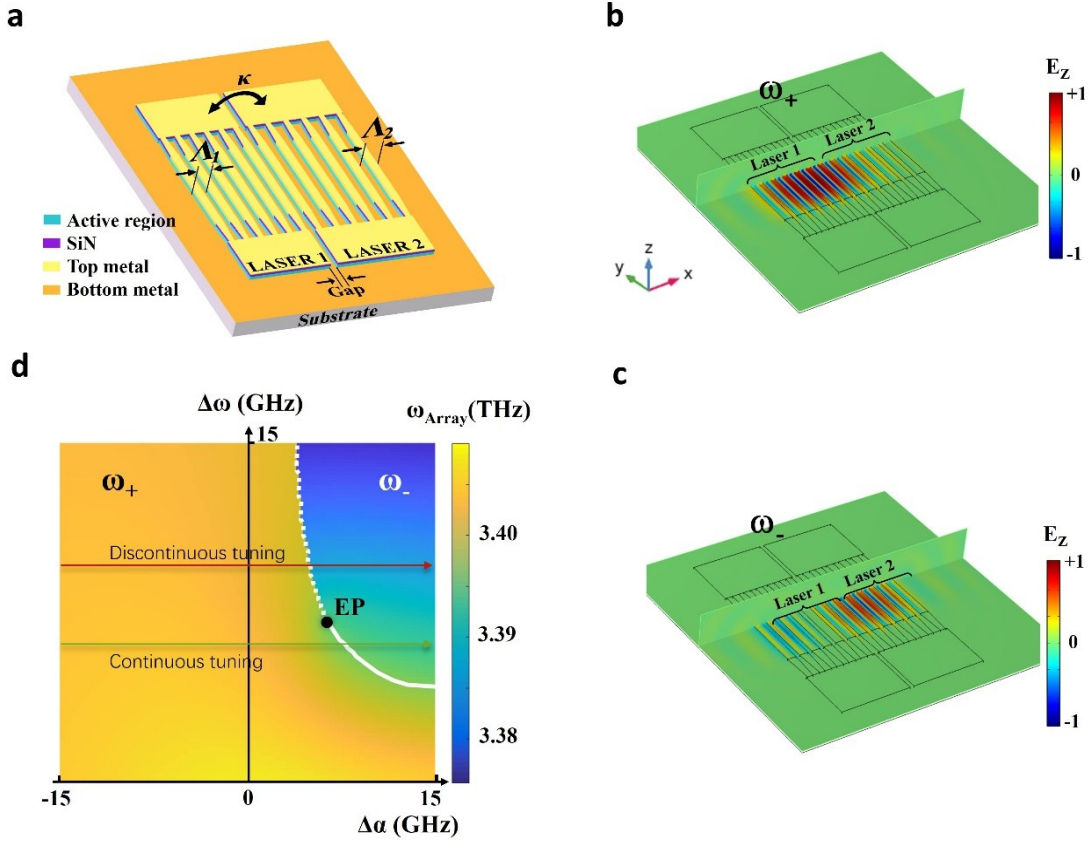

**Figure 2 | Numerical simulations of frequency tuning in a dual-THz-QCL array. a,** Schematic of two complex-coupled second-order distributed-feedback (DFB) THz QCLs differing only in grating period. The gap between the two sub-lasers acts as a low-index region enabling efficient field exchange. **b, c,** Simulated normalized electric-field ($E_z$) profiles of the two supermodes $\omega_+$ (**b**) and $\omega_-$ (**c**). The calculated values are $\omega_1$ = 3.3965+0.0110i THz, $\omega_2$ = 3.3815+0.0032i THz, $\omega_+$ = 3.3954-0.0016i THz, $\omega_-$ = 3.3854+0.0099i THz, and $\kappa$ = 6.2 – 9.4i GHz, respectively. **d,** Real frequency of the lower-loss supermode mapped in the ($\Delta\omega, \Delta\alpha$) parameter space. The mode-switching boundary consists of a PSB branch (solid white curve), where $\omega_+$ and $\omega_-$ coalesce in real parts but split in imaginary parts, and a PS branch (dashed white curve), where the opposite occurs. The two branches meet at an EP.

Figure 2a schematically illustrates a dual-laser array consisting of two independently pumped second-order distributed-feedback (DFB) THz QCLs operating near 3.4 THz. The grating periods ($\Lambda_1$ and $\Lambda_2$) of two sub-lasers are intentionally detuned to define a frequency detuning $\Delta\omega$, while independent pump currents provide continuous control of the relative loss $\Delta\alpha$ over approximately −15 to +15 GHz. By tailoring the grating geometry, modal extent, and inter-cavity spacing, both the real and imaginary components of $\kappa$ can be engineered. As shown in Supplementary Information S2, $\kappa$ depends moderately on $\Delta\omega$ and $\Delta\alpha$ within the operating range of the array. These design freedoms allow the mode-switching boundary and the EP to be deliberately positioned relative to the accessible tuning trajectories.

Finite-element simulations are used to analyze the supermodes of the arrays. Figures 2b and 2c show representative electric-field ($E_z$) profiles of the $\omega_\pm$ supermodes for an array designed with $\Delta\omega < |\text{Im}(\kappa)|$, such that the tuning trajectories lie below the EP. Additional field evolution and the corresponding Riemann surfaces of $\omega_\pm$ are provided in Supplementary Information S3 and S4. These simulations show that current-induced redistribution of the modal fields between the two sub-lasers constitutes the dominant mechanism for frequency tuning.

To directly link the device behavior to the topology of the mode-switching boundary, we numerically map the real frequency of the lower-loss supermode $\omega_{\text{Array}}$ over the $(\Delta\omega, \Delta\alpha)$ space, as shown in Fig. 2d. The PS and PSB branches of the boundary are clearly resolved and meet at the EP, in excellent agreement with the general theoretical model (Fig. 1d). Tuning trajectories above the EP intersect the PS branch and result in discontinuous hopping of the lasing frequency, whereas trajectories below the EP cross the PSB branch and support continuous frequency tuning. Numerical simulations (Supplementary Information S5) further show that, for a given variation range of $\Delta\alpha$ (from -15 to 15 GHz), the continuous tuning range increases with $\Delta\omega$ and is maximized when the trajectory passes through the EP. Because forcing such a trajectory is experimentally challenging, a practical strategy is to realize robust continuous tuning by operating in the regime below the EP.

The close quantitative agreement between the numerical simulations of the specific THz QCL structures and the general two-level non-Hermitian model – despite the simplifying assumption of constant $\kappa$ in the theory – reveals that the mode-switching boundary and its topology are intrinsic features of complex-coupled laser systems.

**Experimental frequency tuning in non-Hermitian dual-THz-QCL arrays**

To verify the theoretical predictions, we fabricated two dual-THz-QCL arrays to demonstrate both continuous and discontinuous frequency tuning. In Array 1, the grating periods are $\Lambda_1 = 35$ μm and $\Lambda_2 = 37$ μm. The calculated $\Delta\omega = 7.1$ GHz and $\kappa \approx 4.5\text{–}9.3i$ GHz, corresponding to $\Delta\omega < |\text{Im}(\kappa)|$, place the tuning trajectory below the EP. In Array 2, $\Lambda_2$ is increased to 38.5 μm, giving $\Delta\omega = 12.3$ GHz and a tuning trajectory above the EP ($\Delta\omega > |\text{Im}(\kappa)|$). The material and device parameters, as well as fabrication approaches, are given in Methods. The inset of Fig. 3a shows a SEM image

of Array 1. Notably, all devices operate entirely without any moving parts.

The tuning trajectory is defined by gradually varying the pump currents of the two sub-lasers ($I_1$ and $I_2$): starting with $I_1 = I_{1,\max}$ and $I_2 = 0$, then increasing $I_2$ to $I_{2,\max}$ at fixed $I_1 = I_{1,\max}$, and finally reducing $I_1$ to zero at fixed $I_2 = I_{2,\max}$. In this way, $\Delta\alpha$ is scanned from a negative maximum through zero to a positive maximum. A normalized parameter $d = \frac{1}{2}[(I_{1,\max} - I_1)/I_{1,\max} + I_2/I_{2,\max}]$ is introduced to represent the pump condition, increasing monotonically from 0 to 1 along the tuning trajectory. Note that $\Delta\alpha$ is not necessarily linearly proportional to $d$, since the loss or gain does not vary linearly or even monotonically with current[45]. Lasing spectrum and output power were recorded at each value of $d$. Measurement details are given in Methods.

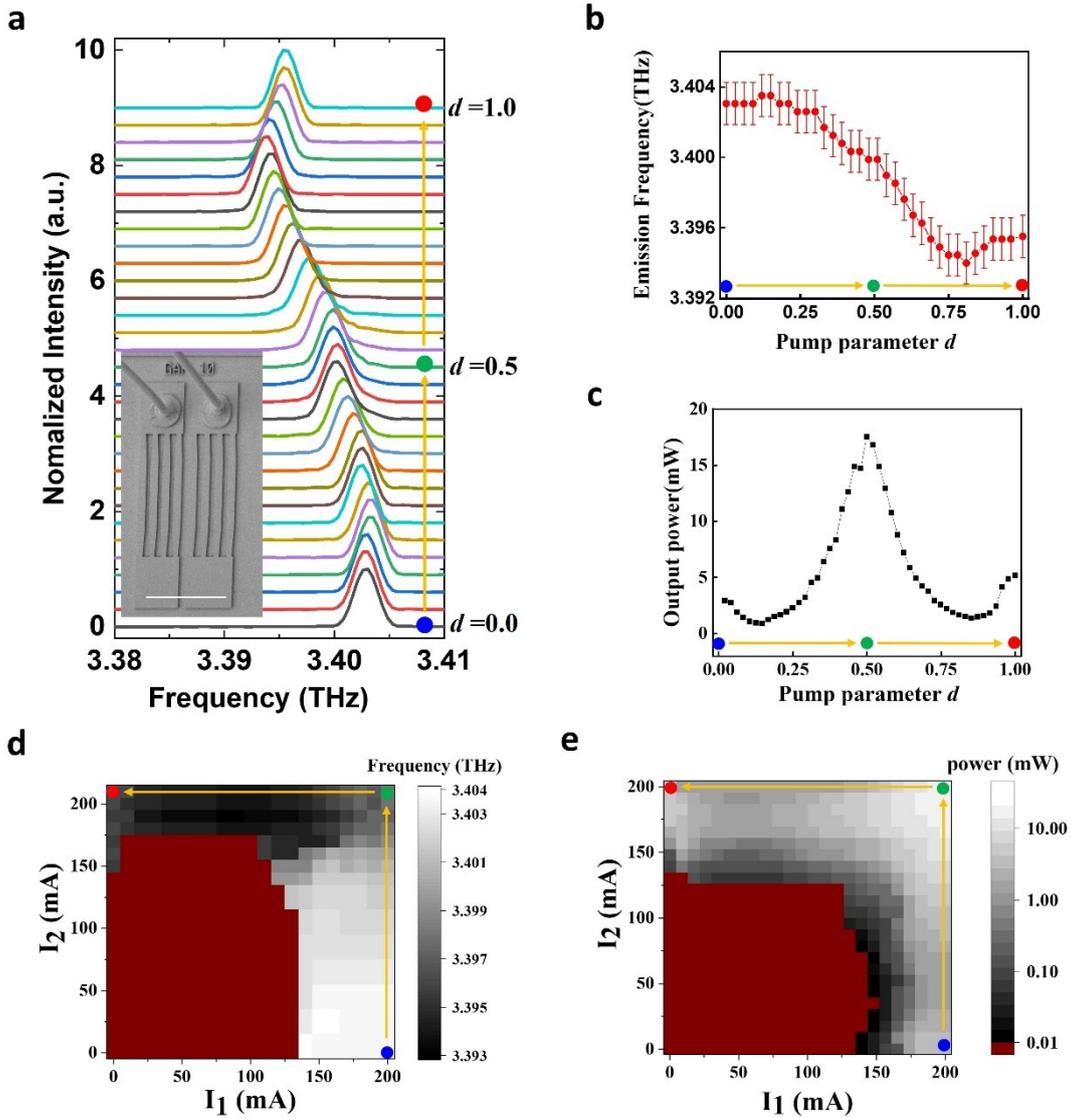

**Figure 3 | Continuous frequency tuning in a dual-THz-QCL array with tuning trajectories**

**below the EP (Array 1). a,** Emission spectra as the tuning parameter *d* increases from 0 to 1 (blue, green, and red dots correspond to *d* = 0, 0.5, and 1). Inset: SEM image of Array 1 (scale bar, 250 μm). **b,c,** Lasing frequency (**b**) and output power (**c**) along the tuning trajectory. Error bars (±1.2 GHz) indicate the spectrometer resolution. **d,e,** Lasing frequency (**d**) and output power (**e**) measured over all combinations of pump currents ($I_1$, $I_2$). Dark red regions indicate no detectable data during spectra and power measurements. Each pixel corresponds to a single measurement without interpolation.

For Array 1 ($\Delta\omega < -\text{Im}(\kappa)$), Fig. 3a presents the emission spectra at different values of *d*. The single-mode emission frequency tunes continuously from 3.404 to 3.394 THz, yielding a tuning range of 10 GHz. This range is slightly smaller than the calculated $2\Delta\omega$ (14.2 GHz), primarily due to the experimentally accessible variation range of $\Delta\alpha$. Figures 3b and 3c show the measured lasing frequency and output power as a function of *d*. The frequency evolution along this trajectory is not strictly monotonic, with extrema appearing at $d \approx 0.2$ and 0.8, near which the output power reaches local minima. The evolution of lasing frequency is qualitatively consistent with the theoretical model based on linear gain. A quantitative description of the frequency and power evolution requires precise loss-gain dependence on pump currents, and a fully nonlinear Maxwell-Bloch treatment accounting for gain saturation above threshold[46]. Nevertheless, the continuous frequency tuning from near $\omega_1$ to near $\omega_2$ along the tuning trajectory is unambiguous, confirming the predicted continuous tuning behavior below the EP.

When the two sub-lasers are uniformly biased, such that the array effectively behaves as a Hermitian system embedded in a uniform gain or loss background, the emission spectra (Supplementary Information S6) exhibit a total frequency shift less than 1.0 GHz from the threshold to the negative differential resistance regime, arising from thermal or carrier effects. This comparison demonstrates that the observed 10 GHz tuning range originates from mutual coupling and is a direct consequence of the non-Hermitian nature of the array. From a topological perspective, the continuous frequency evolution observed here reflects the fact that the experimentally accessible tuning trajectory crosses the PSB branch of the mode-switching boundary, which allows the lower-loss supermode to evolve smoothly between the two Riemann surfaces without frequency hopping.

Figure 3d presents a non-interpolated grayscale map of the lasing frequency as a function of the pump currents ($I_1$, $I_2$). The results show that the complex-coupled laser array supports multiple pump conditions yielding identical emission frequencies, and thus arbitrarily many combinations of ($I_1$, $I_2$) for continuous frequency tuning. This behavior is in fundamental contrast to conventional external-cavity lasers, which support only a single tuning trajectory. Strikingly, continuous frequency tuning persists even when the array operates well above threshold. Figure 3e shows the corresponding output power map, and together with Fig. 3d enables identification of monotonic tuning trajectories with reduced power variation, which are desirable for practical applications.

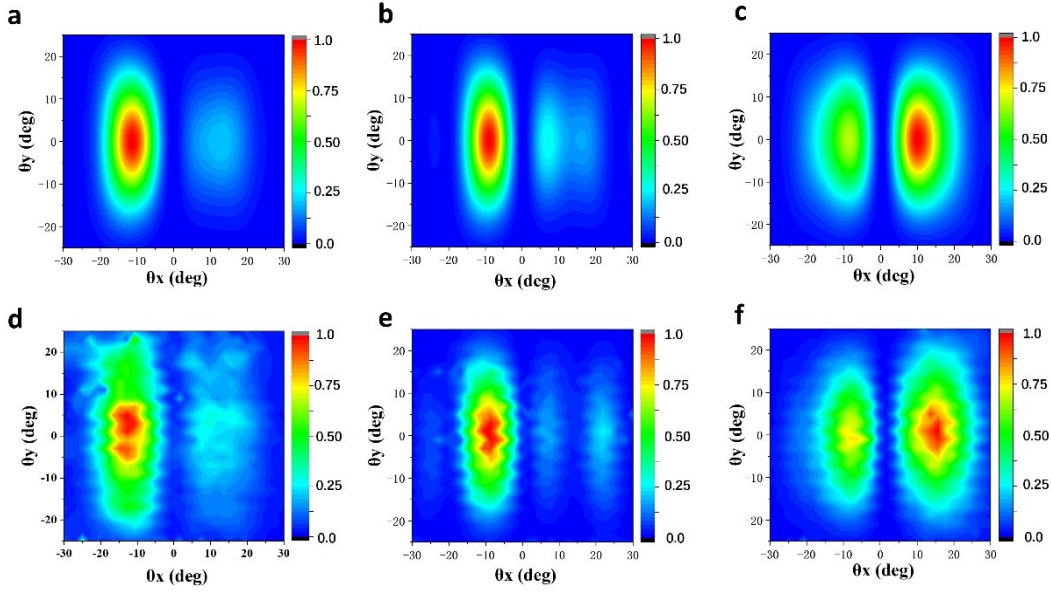

**Figure 4 | Simulated and measured far-field beam patterns of Array 1 under different pump conditions. a–c,** Simulated far-field patterns for representative relative losses Δα = −15, 0, and 15 GHz. **d–f,** Measured far-field beam patterns when the pump parameter *d* equals to 0, 0.5, and 1, respectively.

Figure 4 shows the simulated and measured far-field beam patterns of Array 1 at different pumping conditions (*d* = 0, 0.5, and 1.0, respectively). The measured patterns agree excellently with simulations, confirming that lasing originates from the coupled-array supermodes rather than from individual sub-lasers. The beam patterns exhibit the characteristic features of second-order DFB lasers, namely, two lobes along the grating axis with a central intensity null. As the pump currents vary, the redistribution of the field profiles leads to corresponding changes in the relative lobe profiles.

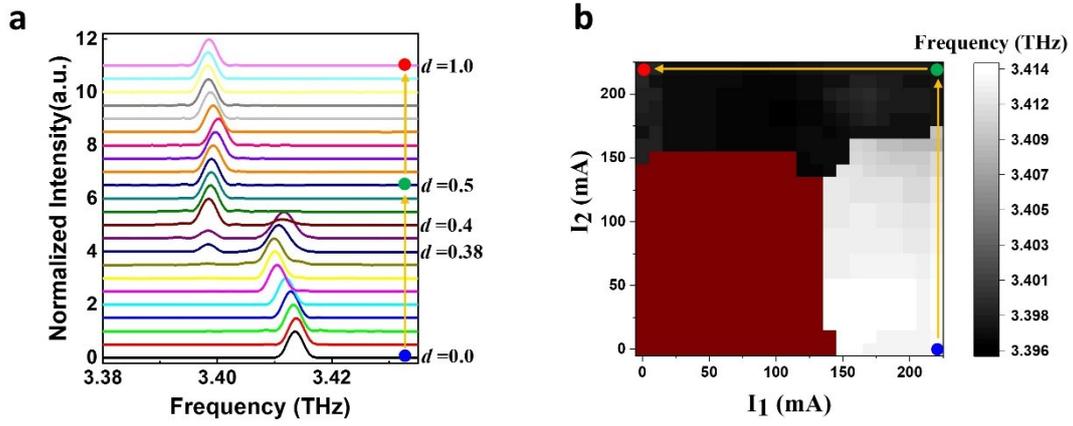

**Figure 5 | Discontinuous frequency tuning in a dual-THz-QCL array with tuning trajectories above the EP (Array 2). a,** Emission spectra as the tuning parameter $d$ varies from 0 to 1 (blue, green, and red dots correspond to $d$ = 0, 0.5, and 1, respectively). **b,** Lasing frequency measured over all combinations of pumping currents ($I_1$, $I_2$). The dark red region indicates below-threshold operation. Each pixel represents a single measurement without interpolation.

For Array 2 ($\Delta\omega > |\text{Im}(\kappa)|$), the lasing behavior is qualitatively different. As shown in Fig. 5a, the array initially operates on the $\omega_+$ supermode for $d \leq 0.35$, with the lasing frequency varying continuously with $d$. In the range $d \approx 0.40$, the system resides close to the PS branch, where $\omega_+$ and $\omega_-$ are split in their real parts but remain close in their imaginary parts, resulting in the simultaneous observation of both supermodes. For $d \geq 0.45$, lasing switches to the $\omega_-$ supermode, demonstrating a discontinuous frequency hopping. Figure 5b presents a non-interpolated grayscale map of the lasing frequency over all combinations of $I_1$ and $I_2$ (with only the dominant mode recorded in the case of dual-mode emission), clearly showing that continuous tuning from near $\omega_1$ to near $\omega_2$ is not achievable along any tuning trajectory. This sharp contrast with Array 1 directly manifests the topology of the mode-switching boundary in parameter space: trajectories intersecting the PS branch are topologically forced to undergo discontinuous frequency hopping, whereas only trajectories travelling through the PSB branch can support continuous frequency tuning.

**Broadband continuous frequency tuning in a non-Hermitian multi-laser array**

The concept was further extended to multi-laser arrays to realize broadband continuous frequency tuning. Exploiting the compactness and integrability of semiconductor lasers, we monolithically integrated 22 sub-lasers into a linear array. Figures 6a–b show the schematic and SEM image of the device; detailed structural and material parameters are given in Methods. Optically, the array can be regarded as a chirped grating with periods ranging from 40.0 to 63.1 μm, while electrically it is divided into independently biased sub-lasers. The designed intrinsic frequencies of the sub-lasers decrease monotonically from 3.520 to 3.360 THz, with frequency detuning between adjacent elements of 3.1–5.2 GHz and designed coupling coefficients ranging from 25.2–8.0i to 19.7–10.7i GHz.

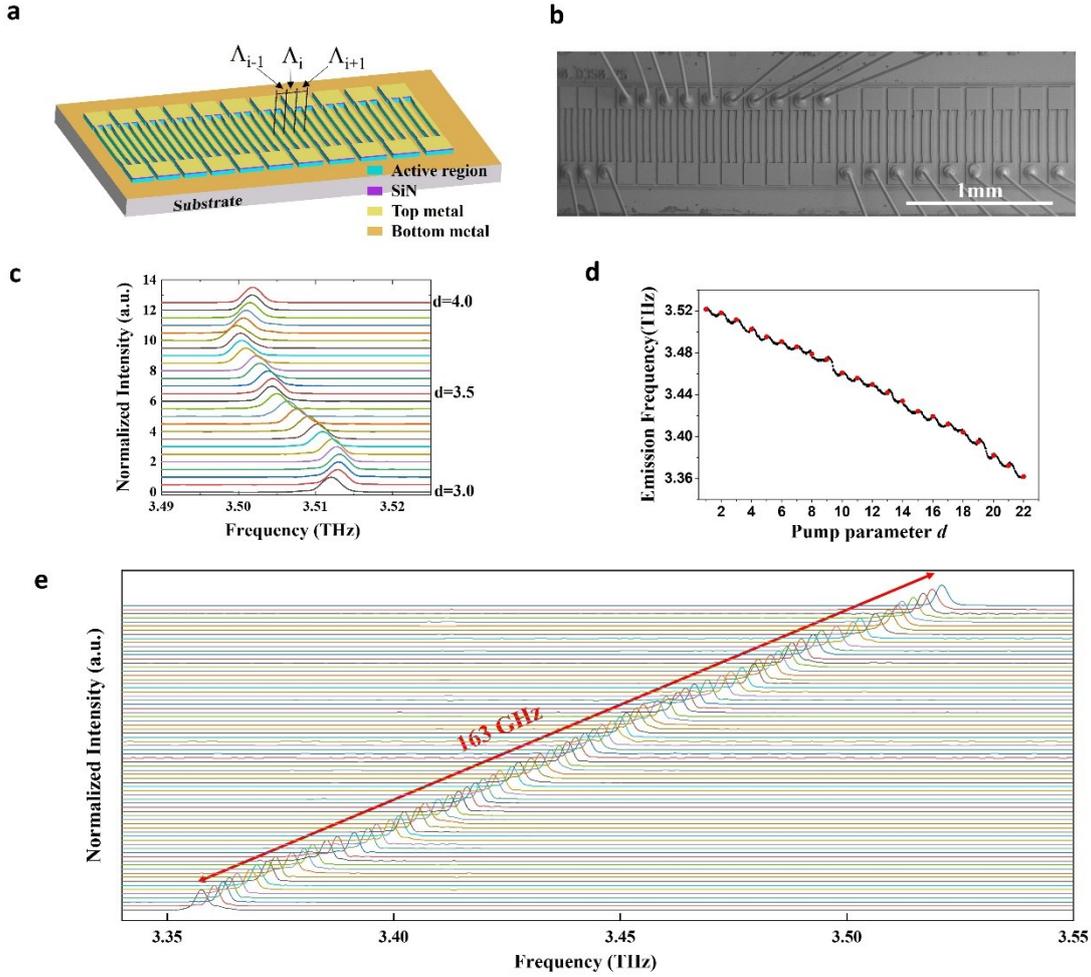

**Figure 6 | Broadband continuous frequency tuning in a non-Hermitian THz-QCL multi-laser array. a, b,** Schematic and SEM image of a 22-sub-laser THz-QCL array with their grating periods vary from 40.0 μm to 63.1 μm. **c,** Emission spectra as the tuning parameter $d$ increases from 3 to 4, showing continuous frequency tuning between two adjacent sub-lasers. **d,** Evolution of the lasing frequency as $d$ increases from 1 to 22. Red points indicate bias conditions where only the $i$th sub-laser ($d = i$) is fully pumped and all others are inactive. **e,** Emission spectra at selected bias conditions, confirming single-mode operation with a SMSR of ~20 dB and a total continuous tuning range of 163 GHz.

To translate the topology-guided tuning principle from dual-laser systems to a multi-element platform, the broadband tuning range is divided into a sequence of tuning segments, each realized by selectively pumping only two adjacent sub-lasers at a time. Specifically, during each tuning segment, only the *i*-th and (*i*+1)-th sub-lasers are electrically pumped, while all other elements remain inactive. This strategy effectively localizes the extended array into an equivalent two-level non-Hermitian subsystem. As a result, the tuning dynamics within each segment remain fully governed by the topology of the mode-switching boundary established for dual-laser systems. Importantly, this localization suppresses the formation of multiple EPs and avoids uncontrolled multimode competition.

Following the dual-laser tuning protocol, the pump parameter is generalized as $d = i + \frac{1}{2}[(I_{i,\text{Max}} - I_i)/I_{i,\text{Max}} + I_{i+1}/I_{i+1,\text{Max}}]$, with $d \in [i, i+1]$. As an example, Fig. 6c shows the emission spectra as *d* varies from 3.0 to 4.0, where the lasing frequency tunes continuously over 12.6 GHz. The tuning is not strictly monotonic, similar to the case of a dual-laser array (Fig. 3a), probably due to the current–gain relations and the coupling to neighboring inactive sub-lasers.

The full tuning behavior of the 22-element array is summarized in Fig. 6d, where lasing frequencies are recorded over 551 pump conditions. As *d* increases from 1 to 22, the emission frequency evolves smoothly from 3.359 to 3.522 THz, demonstrating broadband continuous tuning across the array. The red points in Fig. 6d correspond to bias conditions where only a single sub-laser is fully pumped, indicating that the total tuning span closely matches the intrinsic frequency range defined by the first and last elements of the array. Notably, the presence of multiple admissible tuning trajectories enables pump selection for monotonic frequency evolution, reflecting the existence of topologically equivalent trajectories within the continuous-tuning regime.

After selecting appropriate pump conditions, Fig. 6e demonstrates a monotonic continuous tuning range of 163 GHz, without any moving parts and comparable to state-of-the-art external-cavity systems[47]. Throughout the tuning process, single-mode emission is maintained with a side-mode suppression ratio of approximately 20 dB. The achievable tuning range can be further expanded by increasing the number of sub-lasers or by optimizing the frequency detuning Δω and coupling coefficient κ, with the ultimate limit imposed solely by the gain bandwidth of the active material.

**Conclusion**

We identify a general physical principle for frequency tuning in two-level non-Hermitian systems, governed by the topology of a mode-switching boundary in parameter space. The continuity, discontinuity, and maximal range of frequency tuning are determined solely by how tuning trajectories intersect this boundary and its exceptional point. This topological framework is independent of specific laser platforms and provides a universal route for dynamic eigenvalue engineering in non-Hermitian photonics and related wave systems.

**Method**

**Materials.** The active region employed is based on a bound-to-continuum active transition with a one-well injector, similar to the design reported in Ref.48. The layer sequence starting from the injector barrier is **5.5**/11.0/**1.8**/11.5/**3.8**/<u>9.4</u>/**4.2**/18.4 (in nm), where $Al_{0.15}Ga_{0.85}As$ layers are indicated in bold, GaAs layers in Roman, and the underlined number corresponds to the layer with a Si doping concentration of $2 \times 10^{16}$ cm$^{-3}$. The active region consists of 180 stages with a total thickness of 11.8 μm.

**Device parameters of dual-laser arrays.** Each sub-laser comprises a 6-period DFB grating, with high-index regions formed by 24-μm-wide, 200-μm-long straight ridges. In the low-index regions, both top metallization and active material were removed via lift-off and dry etching. Each ridge is terminated by 80-μm tapers narrowing from 24 to 15 μm, followed by pad areas connecting the ridges for current injection. All non-ridge regions are unpumped by inserting a SiN layer beneath the top metallization. The lasing frequency of each sub-laser is controlled by adjusting the grating period, while the inter-laser gap – matching the ridge pitch – ensures efficient field coupling.

**Device parameters of multi-laser arrays.** Each ridge follows the same structure as in the dual-laser arrays, but each sub-laser incorporates only 3 grating periods. The full array integrates 22 sub-lasers, totaling 66 ridges. The spacing between adjacent ridges – whether within a sub-laser or between neighboring sub-lasers – gradually increases from 16.0 μm to 39.1 μm in 0.35 μm steps.

**Fabrication.** The device is based on metal-metal waveguide, and the processes of fabrication are similar to those described in Ref. 49. After epitaxy, the sample was bonded on an n$^+$ GaAs substrate with an Au−Au interface. The semi-insulator GaAs substrate was then removed via mechanical polishing followed by selective wet etching. After the GaAs/AlGaAs epitaxy was exposed, the top n$^+$ GaAs contact layer was completely removed by wet etching. After that, unpumped regions are defined by deposition of a SiN layer, photolithography, and dry etching. The top metallization (Ti/Au/Ni), which serves as both grating pattern and electrode, was then formed on the top of the active region, defined by contact photolithography, e-beam evaporation, and lift-off. Later, using the top metallization as the self-aligned mask, the low-index region in the grating, as well as the gap between the sub-lasers were formed by removing the whole epilayers via chlorine-based inductively coupled plasma (ICP) etching. Finally, the back-side process consisted of substrate thinning, Ti/Au evaporation, and sample cutting.

**Measurements.** All measurements were performed at a heat-sink temperature of 20 K. Spectral characteristics were obtained using a Fourier transform infrared spectrometer (Bruker 80V) with 2.4 GHz resolution. Data in the main text were measured in pulsed mode (1 μs pulse width, 10 kHz repetition). For typical dual-laser arrays, emission spectra were recorded in both pulsed and continuous-wave modes, showing no linewidth difference, confirming that the linewidth is spectrometer-limited. Pulsed output power was measured with a calibrated Golay detector. Far-field beam patterns were mapped by scanning the Golay detector on a 15-cm-radius sphere centered on the device.

## Acknowledgement

This work was supported by the Strategic Priority Research Program of the Chinese Academy of Sciences (Nos. XDB0980000), and the National Natural Science Foundation of China (Nos. 12393833, 62435020, 62235010, and 12274285).

# Broadband Continuous Frequency Tuning in Non-Hermitian Laser Arrays Enabled by Mode-Switching Boundary Topology

Chuanfeng Yan[1,2,†], Cheng Tan[1,2,†], Kai Wang[1], Hongzhou Bai[1,2], Shanhai Gao[1,2], Lianghua Gan[1], Yueheng Zhang[3,*], Qijie Wang[4], Gangyi Xu[1,*]

5. *National Key Laboratory of Infrared Detection Technologies, Shanghai Institute of Technical Physics, Chinese Academy of Sciences, Shanghai 200083, China*
6. *University of Chinese Academy of Sciences, Beijing 100049, China*
7. *Key Laboratory of Artificial Structures and Quantum Control, School of Physics and Astronomy, Shanghai Jiao Tong University, Shanghai 200240, China*
8. *Division of Physics and Applied Physics, School of Physical and Mathematical Sciences, Nanyang Technological University, 21 Nanyang Link, Singapore 637371, Singapore*

\* Corresponding authors: *yuehzhang@sjtu.edu.cn, gangyi.xu@mail.sitp.ac.cn*

† These authors contribute equally to this work.

## Content

S1. Mathematical derivation of frequency tuning in two-level non-Hermitian laser arrays.

S2. Evolution of the complex coupling coefficient in dual-THz-QCL arrays.

S3. Field profile of a dual-THz-QCL array under different pump currents.

S4. Numerical calculations of the complex eigenfrequencies of the supermodes in the dual-THz-QCL arrays.

S5. Numerical calculations of the tuning ranges of the dual-THz-QCL arrays.

S6. Emission spectra of the dual-THz-QCL array (Array1) with uniform pump.

**S1. Mathematical derivation of frequency tuning in two-level non-Hermitian laser arrays.**

The complex expression for the electric field of electromagnetic wave can be written as

$$E(t) = A exp(i(\omega + \alpha i)t) \tag{S1}$$

where ω is the frequency, α is either loss (positive) or gain (negative), and $A$ is the amplitude. In a complex-coupled dual-laser array, the time evolution of electric field ($E_1$ and $E_2$) in sub-lasers 1 and 2 can be modeled by the temporal coupled-mode theory (TCMT) and be described as

$$\frac{dE_1(t)}{dt} = i(\omega_1 + \alpha_1 i)E_1(t) + ik_{12}E_2(t) \tag{S2}$$

$$\frac{dE_2(t)}{dt} = i(\omega_2 + \alpha_2 i)E_2(t) + ik_{21}E_1(t) \tag{S3}$$

Here, $\omega_i$ and $\alpha_i$ (i=1, 2) are the intrinsic frequency and net loss rate of the $i$-th sub-laser before coupling, respectively. $\kappa_{12}$ and $\kappa_{21}$ are the complex-valued coupling coefficients between the two sub-lasers. For a reciprocal coupled system, $\kappa_{12}$ equals to $\kappa_{21}$, and can be defined as:

$$\kappa_{12} = \kappa_{21} = \kappa = Re(\kappa) + iIm(\kappa) \tag{S4}$$

Therefore, the dual-laser system can be described in terms of Hamiltonian H:

$$\frac{d}{dt}\begin{pmatrix}E_1\\E_2\end{pmatrix} = iH\begin{pmatrix}E_1\\E_2\end{pmatrix} \tag{S5}$$

$$H = \begin{pmatrix}\Delta\omega + \Delta\alpha i & \kappa \\ \kappa & -\Delta\omega - \Delta\alpha i\end{pmatrix} + (\omega_0 + \alpha_0 i)\begin{pmatrix}1 & 0\\0 & 1\end{pmatrix} = \widetilde{H} + (\omega_0 + \alpha_0 i)I \tag{S6}$$

Here, $\omega_0 = (\omega_1 + \omega_2)/2$ and $\alpha_0 = (\alpha_1 + \alpha_2)/2$ are the average frequency and loss of the system. $\Delta\omega = (\omega_1 - \omega_2)/2$ and $\Delta\alpha = (\alpha_1 - \alpha_2)/2$ represent the frequency detuning and relative loss of the two sub-lasers, respectively. As stated in the main text, $\omega_1 > \omega_2$, therefore $\Delta\omega > 0$. $I$ is the identity matrix, and $\widetilde{H}$ represents the effective Hamiltonian of the system. The eigenvalues of $H$, denoted as $\omega_\pm$, can be expressed as:

$$\omega_\pm = (\omega_0 + i\alpha_0) \pm \sqrt{(\Delta\omega + i\Delta\alpha)^2 + (Re(\kappa) + iIm(\kappa))^2}$$

$$=(\omega_0 + i\alpha_0) \pm \sqrt{\Delta\omega^2 - \Delta\alpha^2 + Re(\kappa)^2 - Im(\kappa)^2 + 2[Re(\kappa)Im(\kappa) + \Delta\omega\Delta\alpha]i} \quad (S7)$$

$$\omega_+ = (\omega_0 + i\alpha_0) + \sqrt{\Delta\omega^2 - \Delta\alpha^2 + Re(\kappa)^2 - Im(\kappa)^2 + 2[Re(\kappa)Im(\kappa) + \Delta\omega\Delta\alpha]i} \quad (S8)$$

$$\omega_- = (\omega_0 + i\alpha_0) - \sqrt{\Delta\omega^2 - \Delta\alpha^2 + Re(\kappa)^2 - Im(\kappa)^2 + 2[Re(\kappa)Im(\kappa) + \Delta\omega\Delta\alpha]i} \quad (S9)$$

Note, the array operates on the lower-loss supermode, which has a smaller imaginary part of the complex eigenfrequency. We define a complex value ($x+iy$) that satisfies

$$x + iy = \sqrt{\Delta\omega^2 - \Delta\alpha^2 + Re(\kappa)^2 - Im(\kappa)^2 + 2[Re(\kappa)Im(\kappa) + \Delta\omega\Delta\alpha]i} \quad (S10)$$

Here $x$, $y$ are real values and $x \geq 0$. If $Re(\kappa)Im(\kappa) + \Delta\omega\Delta\alpha < 0$, $x$ and $y$ are opposite in sign, and $\omega_+$ mode is the lower-loss mode. In contrast, if $Re(\kappa)Im(\kappa) + \Delta\omega\Delta\alpha > 0$, $x$ and $y$ are the same in sign, and $\omega_-$ mode is the lower-loss mode. Therefore, the system operates on the $\omega_+$ mode when $Re(\kappa)Im(\kappa) + \Delta\omega\Delta\alpha < 0$, and switches to the $\omega_-$ mode when $Re(\kappa)Im(\kappa) + \Delta\omega\Delta\alpha > 0$. So, the mode-switching boundary can be expressed by

$$Re(\kappa)Im(\kappa) + \Delta\omega\Delta\alpha = 0 \quad (S11)$$

At this boundary, the eigenfrequency satisfies the following equations.

$$\omega_\pm = (\omega_0 + i\alpha_0) \pm \sqrt{\Delta\omega^2 - \Delta\alpha^2 + Re(\kappa)^2 - Im(\kappa)^2} \quad (S12)$$

Substitute Eq. (S11) into Eq. (S12), we have

$$\omega_\pm = (\omega_0 + i\alpha_0) \pm \sqrt{[\Delta\omega^2 + Re(\kappa)^2]\left[1 - \left(\frac{Im(\kappa)}{\Delta\omega}\right)^2\right]} \quad (S13)$$

Therefore, on the mode-switching boundary, if $\Delta\omega < |Im(\kappa)|$, the last term in Eq. (S13) is a pure imaginary value, and therefore $\omega_+$ and $\omega_-$ degenerate in the real part but split in the imaginary part, denoting the pseudo-symmetry-broken (PSB) branch. In contrast, if $\Delta\omega > |Im(\kappa)|$, the last term in Eq. (S13) is a pure real value, and therefore $\omega_+$ and $\omega_-$ degenerate in the imaginary part but split in the real part, representing the pseudo-symmetry (PS) branch. Lastly, if $\Delta\omega = -Im(\kappa)$ and $\Delta\alpha = Re(\kappa)$ (or equivalently, $\Delta\omega = Im(\kappa)$ and $\Delta\alpha = -Re(\kappa)$, depending on the sign of Im(κ)), $\omega_+$ and $\omega_-$ degenerate in both the real and imaginary parts, corresponding to an exceptional point (EP) in the parameter space of ($\Delta\omega$, $\Delta\alpha$) that connects the PS and PSB branches.

Now we derive the extrema of the real parts of the eigenfrequencies Re($\omega_\pm$) along all possible tuning trajectories. As we state in the main text, the tuning trajectories are approximately parallel to the $\Delta\alpha$ axis in the ($\Delta\omega$, $\Delta\alpha$) parameter space. From Eq. (S10), we have

$$\omega_+ = (\omega_0 + i\alpha_0) + (x + iy) \qquad (S14a)$$

$$\omega_- = (\omega_0 + i\alpha_0) - (x + iy) \qquad (S14b)$$

$$x^2 - y^2 = \Delta\omega^2 - \Delta\alpha^2 + Re(\kappa)^2 - Im(\kappa)^2 \qquad (S15a)$$

$$xy = Re(\kappa)Im(\kappa) + \Delta\omega\Delta\alpha \qquad (S15b)$$

We take $\omega_+$ as an example, the extremum of Re($\omega_+$) equals to the extremum of $x$. Substituting Eq. (S15b) into Eq. (S15a), we have

$$x^2 = \frac{\Delta\omega^2 - \Delta\alpha^2 + Re(\kappa)^2 - Im(\kappa)^2}{2} + \sqrt{\left(\frac{\Delta\omega^2 - \Delta\alpha^2 + Re(\kappa)^2 - Im(\kappa)^2}{2}\right)^2 + (Re(\kappa)Im(\kappa) + \Delta\omega\Delta\alpha)^2} \quad (S16)$$

To find the extremum of $x$, we let $\frac{dx}{d(\Delta\alpha)} = 0$, which results in

$$\sqrt{(\Delta\omega^2 - \Delta\alpha^2 + Re(\kappa)^2 - Im(\kappa)^2)^2 + 4(Re(\kappa)Im(\kappa) + \Delta\omega\Delta\alpha)^2} =$$

$$2[Re(\kappa)Im(\kappa) + \Delta\omega\Delta\alpha]\Delta\omega/\Delta\alpha - (\Delta\omega^2 - \Delta\alpha^2 + Re(\kappa)^2 - Im(\kappa)^2) \qquad (S17)$$

Here, we assume that the coupling coefficient $\kappa$ is independent on $\Delta\alpha$, $i.e.$, $\frac{dRe(\kappa)}{d(\Delta\alpha)} = \frac{dIm(\kappa)}{d(\Delta\alpha)} = 0$. Square both sides of Eq. (S17), and after some algebraic derivation, we get

$$[Re(\kappa)Im(\kappa) + \Delta\omega\Delta\alpha]^2 = [Re(\kappa)Im(\kappa) + \Delta\omega\Delta\alpha]^2\left(\frac{\Delta\omega}{\Delta\alpha}\right)^2 - [\Delta\omega^2 - \Delta\alpha^2 + Re(\kappa)^2 - Im(\kappa)^2]\left(\frac{\Delta\omega}{\Delta\alpha}\right)[Re(\kappa)Im(\kappa) + \Delta\omega\Delta\alpha] \qquad (S18)$$

After some algebraic derivation, we get

$$[\Delta\omega^2 - \Delta\alpha^2][Re(\kappa)Im(\kappa) + \Delta\omega\Delta\alpha] = [\Delta\omega^2 - \Delta\alpha^2 + Re(\kappa)^2 - Im(\kappa)^2]\Delta\omega\Delta\alpha \qquad (S19)$$

Equation (S19) can be rewritten as

$$\left[\frac{\Delta\omega}{\Delta\alpha}\frac{Re(\kappa)}{Im(\kappa)}+1\right]\left[\frac{\Delta\omega}{\Delta\alpha}-\frac{Re(\kappa)}{Im(\kappa)}\right]=0 \tag{S20}$$

Therefore, the extrema of $x$ (and thus $Re(\omega_\pm)$) and the related values of $\Delta\alpha$ can be expressed as the following.

Case 1: when $\Delta\alpha = \frac{\Delta\omega Im(\kappa)}{Re(\kappa)}$,

$$x_{max} = \sqrt{Re(\kappa)^2 + \Delta\omega^2},$$

$$Re(\omega_+)_{max} = \omega_0 + \sqrt{Re(\kappa)^2 + \Delta\omega^2},$$

$$Re(\omega_-)_{min} = \omega_0 - \sqrt{Re(\kappa)^2 + \Delta\omega^2} \tag{S21a}$$

Case 2: when $\Delta\omega \geq |Im(\kappa)|$ and $\Delta\alpha = -\frac{\Delta\omega Re(\kappa)}{Im(\kappa)}$,

$$x_{min} = \sqrt{\Delta\omega^2 - Im(\kappa)^2},$$

$$Re(\omega_+)_{min} = \omega_0 + \sqrt{\Delta\omega^2 - Im(\kappa)^2},$$

$$Re(\omega_-)_{max} = \omega_0 - \sqrt{\Delta\omega^2 - Im(\kappa)^2}, \tag{S21b}$$

Case 3: when $\Delta\omega \leq |Im(\kappa)|$, and $\Delta\alpha = -\frac{Re(\kappa)Im(\kappa)}{\Delta\omega}$ (the condition of mode-switching boundary),

$$x_{min} = 0, Re(\omega_+)_{min} = Re(\omega_-)_{max} = \omega_0 \tag{S21c}$$

Case 4: when $\Delta\alpha \to \pm\infty$, $\omega_+ \to \omega_1$, $\omega_- \to \omega_2$. \qquad (S22d)

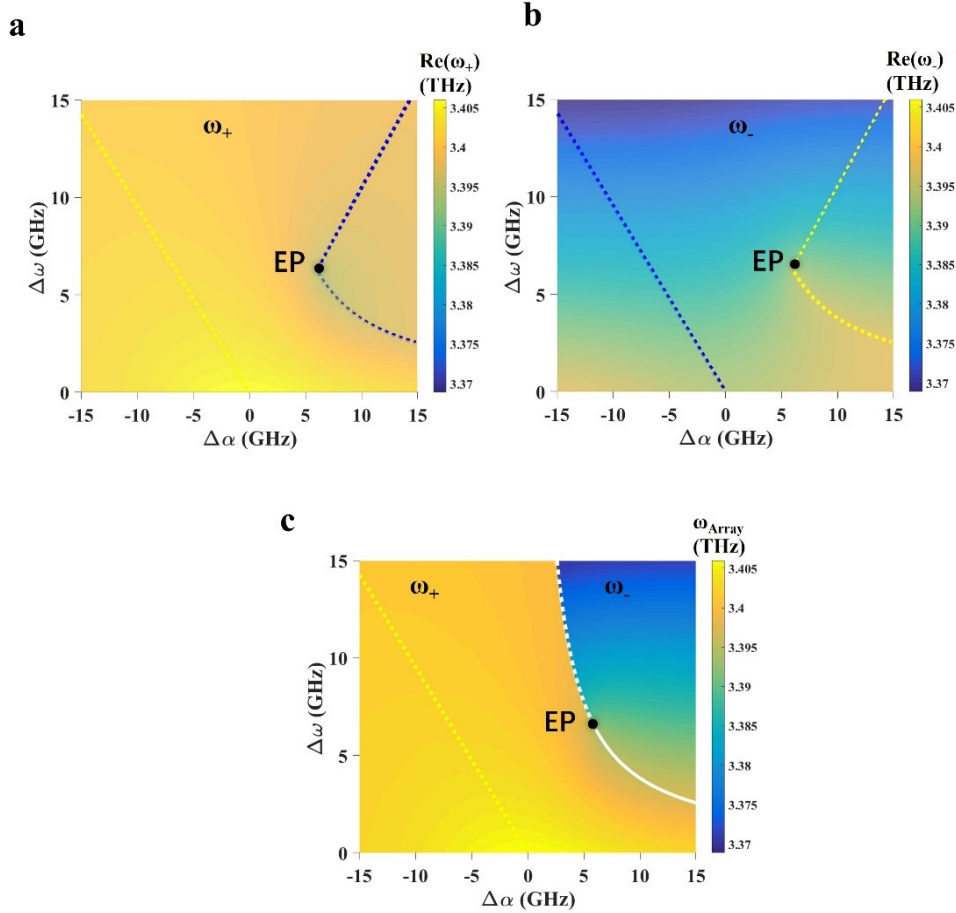

Figure S1 | Evolution of the real frequency of $\omega_+$ (**a**), $\omega_-$ (**b**), and the lower-loss supermode $\omega_{Array}$ (**c**) in the ($\Delta\omega$, $\Delta\alpha$) parameter space. In panels **a** and **b**, the yellow curves correspond to the maximum values of Re($\omega_+$) and Re($\omega_-$) along all tuning trajectories, while the blue curves correspond to their minimum values. In panel **c**, white dashed and solid curves are the PS and PSB branches that meet at the EP. The yellow dashed line denotes the maximum values of $\omega_{Array}$ along all tuning trajectories. For the trajectories below the EP, the PSB branch represents the minimum values of Re($\omega_+$) and also the maximum values of Re($\omega_-$).

The above theoretical analysis is illustrated in Figs. S1a and S1b, which show the evolution of Re($\omega_+$) and Re($\omega_-$), respectively, in the ($\Delta\alpha$, $\Delta\omega$) parameter space. In consistent with the main text, $\omega_1+i\alpha_1$ is set as 3.4+0i (THz), $\kappa$ equals to 6.0-6.3i (GHz), and the scan ranges of $\Delta\omega$ and $\Delta\alpha$ are marked in Fig.1S. In Fig. S1a, the yellow straight line corresponds to the condition $\Delta\alpha = \Delta\omega Im(\kappa)/Re(\kappa)$, on which Re($\omega_+$) reaches its maximum value along each tuning trajectory. By contrast, the blue curve marks the minimum value of Re($\omega_+$). Below the exceptional point (EP), defined by $\Delta\alpha = -Re(\kappa)Im(\kappa)/\Delta\omega$, the blue curve coincides with the PSB branch of the mode-switching boundary, where Re($\omega_+$) approaches its minimum value Re($\omega_+$) = $\omega_0$. Above the EP, the minimum of Re($\omega_+$) occurs at $\Delta\alpha = -\Delta\omega Re(\kappa)/Im(\kappa)$, with $Re(\omega_+)_{min} = \omega_0 +$ $\sqrt{\Delta\omega^2 - Im(\kappa)^2}$. Figure S1b is the mirror image of Fig. S1a, reflecting the relation $Re(\omega_+) + Re(\omega_-) = \omega_1 + \omega_1 = constant$. As a result, when $Re(\omega_+)$ approaches to its maximum value, $Re(\omega_-)$ simultaneously approaches to its minimum value, and vice versa.

Based on the above analysis, Fig. S1c plots the evolution of the lower-loss supermode $\omega_{Array}$ in the ($\Delta\alpha$, $\Delta\omega$) parameter space. Figure S1c is consistent with Fig. 1d in the main text, except that an additional yellow dashed line is included, corresponding to $\Delta\alpha = \Delta\omega Im(\kappa)/Re(\kappa)$, which marks the maximum attainable lasing frequency along each tuning trajectory, $\omega_{Array,max} = Re(\omega_+)_{max} = \omega_0 + \sqrt{Re(\kappa)^2 + \Delta\omega^2}$).

We now focus on tuning trajectories below the EP where continuous frequency tuning is achievable. The evolution of the lower-loss supermode $\omega_{Array}$ can be determined as $\Delta\alpha$ is swept from $-\infty$ to $+\infty$. Starting from large negative $\Delta\alpha$, the system operates on the $\omega_+$ supermode, with its real frequency initially close to $\omega_1$. As $\Delta\alpha$ increases, the lasing frequency rises and reaches its maximum value $\omega_{Array,max} = \omega_0 + \sqrt{Re(\kappa)^2 + \Delta\omega^2}$ at $\Delta\alpha = \Delta\omega\, Im(\kappa)/Re(\kappa)$. With further increase of $\Delta\alpha$, the lasing frequency decreases and reaches $\omega_0$ upon approaching the mode-switching boundary. At the boundary, defined by $\Delta\alpha = Re(\kappa)Im(\kappa)/\Delta\omega$, the lasing mode switches from $\omega_+$ to $\omega_-$, with $Re(\omega_+) = Re(\omega_-) = \omega_0$. Upon further increasing $\Delta\alpha$, the system continues to lase on the $\omega_-$ supermode, whose frequency decreases monotonically and asymptotically approaches $\omega_2$ as $\Delta\alpha \rightarrow +\infty$.

Throughout this process, the maximum and minimum lasing frequencies are $\left(\omega_0 + \sqrt{Re(\kappa)^2 + \Delta\omega^2}\right)$ and $\omega_2$, respectively, yielding a total continuous tuning range of $\left(\Delta\omega + \sqrt{Re(\kappa)^2 + \Delta\omega^2}\right)$. The analysis shows that, for a sufficiently large variation of $\Delta\alpha$, the lasing frequency can be continuously tuned from near $\omega_1$ to near $\omega_2$. Notably, for a nonzero $Re(\kappa)$, the achievable tuning range can exceed the intrinsic frequency difference $(\omega_1 - \omega_2)$ of the two uncoupled sub-lasers. Furthermore, the total tuning range is a monotone increasing function of $\Delta\omega$. Considering the condition of continuous tuning, $\Delta\omega \leq |Im(\kappa)|$, the most efficient tuning trajectory is the one passing through the EP, which yields the largest continuous tuning range for a given variation range of $\Delta\alpha$.

## S2. Evolution of the complex coupling coefficient in dual-THz-QCL arrays.

In dual-THz-QCL arrays, we numerically calculated the complex coupling coefficient $\kappa$, and its evolution in the ($\Delta\alpha$, $\Delta\omega$) parameter space. To this aim, we first calculated the complex eigenfrequency ($\omega_i + i\alpha_i$) of the two induvial sub-lasers before coupling. We further calculated the eigenfrequency ($\omega_\pm$) of the coupled array. According to Eq. (S7), we can immediately get the complex value of $\kappa$.

During numerical calculations, we fixed the structural and material parameters of sub-laser 1 and scanned those parameters of sub-laser 2. For sub-laser 1, the grating contains 6 periods, the width of high-index region is 24.0 μm, the period is 35.0 μm, and the net loss of active region is set as zero. For sub-laser 2, the grating also contains 6 periods and the width of high-index region is fixed as 24.0 μm, while the period is scanned from 35.0 to 39.5 μm and the net loss is scanned approximately from 40 to -40 cm$^{-1}$. The calculations of eigenfrequency and field profile were consulted by meanings of finite element method via a commercial package of COMSOL-Multiphysics. The gap between the two sub-lasers is always set to be the average width of the low-index regions of grating in the two sub-lasers. Figures S2a-b show respectively the real and imaginary part of $\kappa$ in the ($\Delta\omega$, $\Delta\alpha$) space. In general, $\kappa$ varies with $\Delta\omega$ and $\Delta\alpha$, but in certain small range of ($\Delta\omega$, $\Delta\alpha$), $\kappa$ can be regarded as a constant.

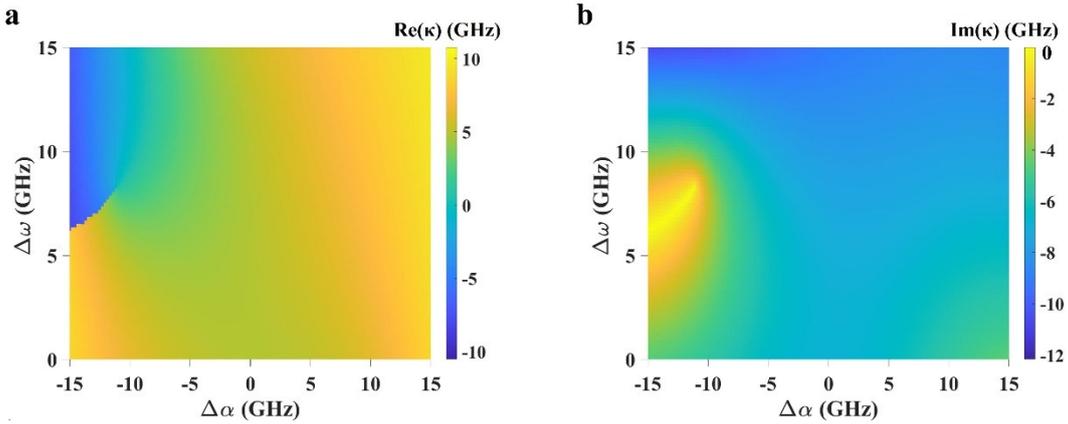

**Figure S2** | **a** and **b** show the evolution of Re($\kappa$) and Im($\kappa$) in the ($\Delta\omega$, $\Delta\alpha$) space of dual-THz-QCL arrays.

## S3. Field profile of a dual-THz-QCL array under different pump currents

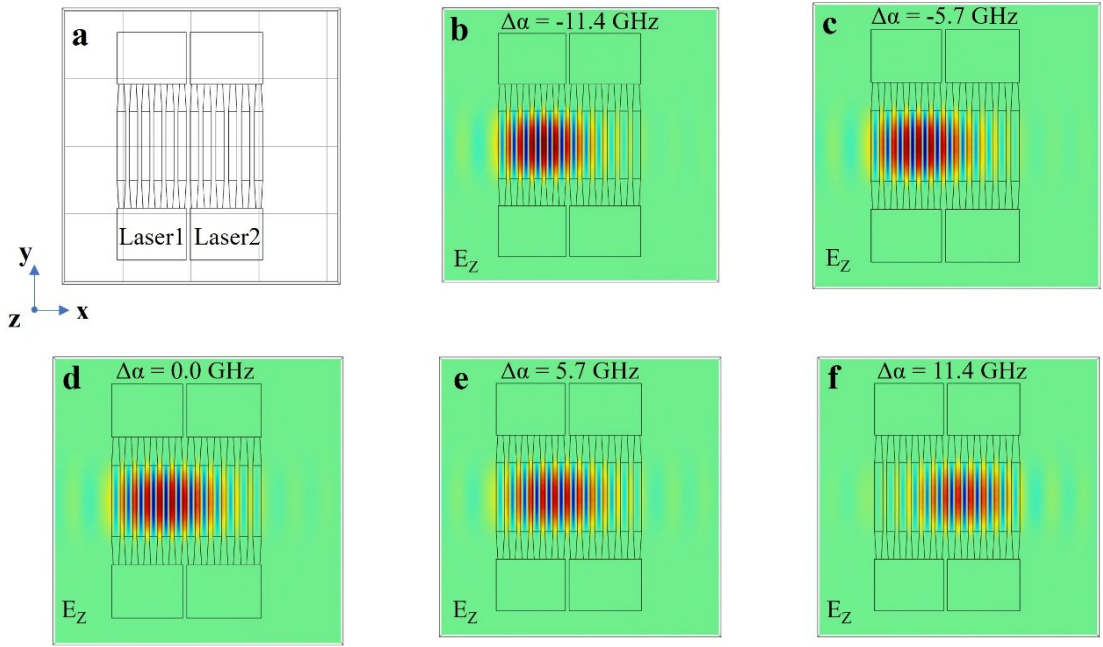

**Figure S3** | **a**, Schematic top view of the coupled dual-THz QCLs. **b–f**, Simulated $E_z$ field profiles of the lower-loss supermode under different relative loss rates $\Delta\alpha$: –11.4, –5.7, 0.0, 5.7, and 11.4 GHz, corresponding to different pump currents applied to sub-lasers 1 and 2.

For an exemplary dual-THz-QCL array, we numerically calculated the field profiles of the lower-loss supermode under different bias conditions, and the results are presented in Figure S3. Figure S3a shows the schematic top view of the array. Both sub-lasers feature DFB grating with 6 periods, and the width of high-index ridge is fixed as 24.0 μm. The only difference between the two sub-lasers is the period, which is 35.0 and 37.0 μm, respectively. The gap between them is 11.0 μm in width, which approximately equals to the width of low-index region of the gratings. $\omega_1$ and $\omega_2$ are calculated to be respectively 3.3965 and 3.3815 THz, while $\alpha_1$ and $\alpha_2$ are variables simulating different pump currents. Figures S3b to S3f illustrate the evolution of filed profiles under different pump currents, i.e., different values of $\Delta\alpha$. Figure S3 reveals it is the redistribution of field profile with the change of pump currents that results in the frequency tuning.

## S4. Numerical calculations of the eigenfrequencies of the supermodes in dual-THz-QCL arrays

During numerical calculations, we fixed the structural and material parameters of sub-laser 1 and scanned those parameters of sub-laser 2. The settings of structural and material parameters are the same as those in section S2.

Figures S4a and S4b present the real and imaginary parts of $\omega_\pm$ in the ($\Delta\omega$, $\Delta\alpha$) parameter space for dual-THz-QCL arrays. Here, the yellow and blue Riemann surfaces correspond to $\omega_+$ and $\omega_-$, respectively. The solid white line is the intersection line of two Riemann surfaces where $\omega_+$ and $\omega_-$ generate in the real parts. In turn, the dashed white line corresponds to the intersection line of two Riemann surfaces where $\omega_+$ and $\omega_-$ generate in the imaginary parts. The solid and dashed white lines meet at an exceptional point (EP) where $\omega_+$ and $\omega_-$ generate simultaneously in both real and imaginary parts. The numerical simulations agree perfectly with the theoretical results shown in Figures 1b and 1c in the main text, although the coupling coefficient $\kappa$ is a variable quantity in the numerical simulations but is approximated as a constant in the theoretical model.

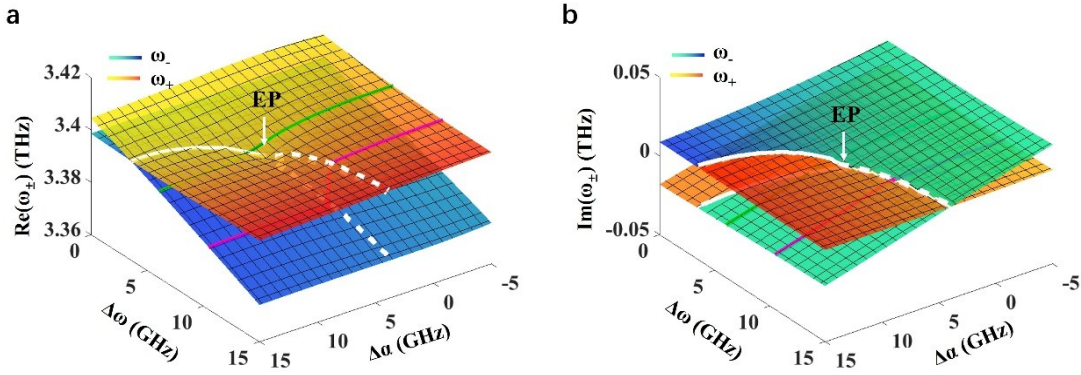

**Figure S4** | Distribution of the real (**a**) and imaginary (**b**) parts of the eigenfrequencies $\omega_+$ and $\omega_-$ obtained from numerical simulations in the ($\Delta\omega$, $\Delta\alpha$) space. The yellow Riemann surface corresponds to $\omega_+$, and the blue one corresponds to $\omega_-$.

## S5. Numerical calculations of the tuning ranges of dual-THz-QCL arrays

Using the same approaches as sections S2 - S4, we fixed the structure and material parameters of sub-laser 1 and scanned the period and net loss of sub-laser 2. In this way, we calculated the real and imaginary parts of the eigenfrequency of the lower-loss supermode in the ($\Delta\omega$, $\Delta\alpha$) space, and the results are shown in Figure S5a and S5b, respectively. Note, Figure S5a is the same as Figure 2d in the main text. The EP occurs when $\Delta\omega \approx$ -Im($\kappa$) $\approx$ 7.2 GHz, and $\Delta\alpha \approx$ Re($\kappa$) $\approx$ 6.3 GHz.

Deduced from Figs. S5a and S5a, Fig. S5c presents the values of $\omega_{Array}$ as a function of $\Delta\alpha$ at different $\Delta\omega$. Clearly, as the tuning trajectory moves from $\Delta\omega$=0 towards EP, the frequency tuning range increases from 2.6 GHz to 13.2 GHz. Above the EP, the frequency tuning becomes discontinuous, perfectly consistent with the theoretical prediction.

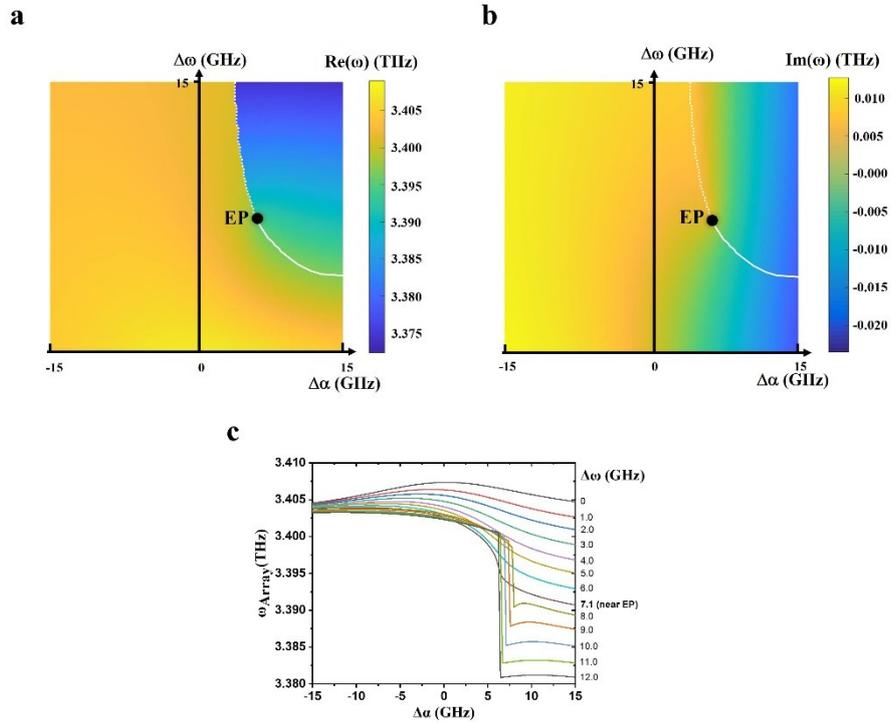

**Figure S5** | **a**, **b**, Real and imaginary parts of the eigenfrequency of the lower-loss supermode in the (Δω, Δα) space. **c**, Evolution of the real frequency of the lower-loss mode ω$_{Array}$ as a function of Δα at different Δω.

## S6. Emission spectra of the dual-THz-QCL array (Array 1) with uniform pump

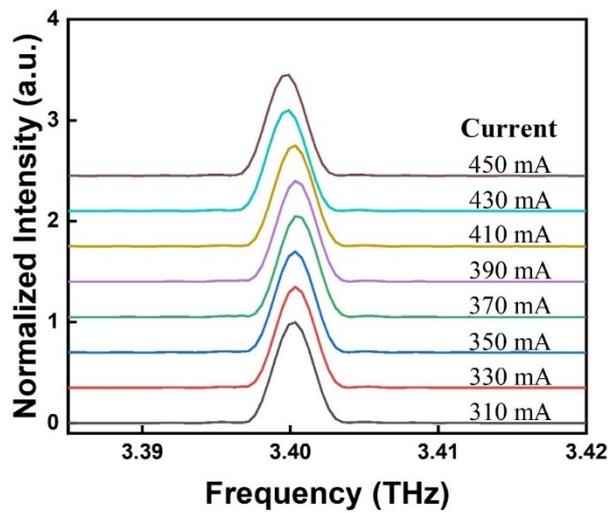

**Figure S6** | Normalized spectra from threshold to peak when the two sub-lasers are equally pumped. The biased currents are marked for each spectrum.

For Array 1 ($\Lambda_1$ = 35 µm, $\Lambda_2$ = 37 µm, $\Delta\omega$ < |Im(K)|), we short-circled the top electrodes of the two sub-lasers and thus uniformly biased them. Under such bias conditions, Figure S6 shows the normalized emission spectra of the array at different pump currents. The array features single mode emission, and the lasing frequency varies less than 1.0 GHz from the threshold (310 mA) to the onset of negative differential resistance (450 mA). The results unambiguously reveal that the frequency tuning caused by thermal effect or the variation of electron density is negligible.